\documentstyle[sprocl,epsf]{article}

\def\gtsim{\mathrel{\raise .5mm \hbox{$>$} 
\kern-2.7mm \lower 1mm \hbox{$\sim$}}}
\def\ltsim{\mathrel{\raise .5mm \hbox{$<$} 
\kern-2.7mm \lower 1mm \hbox{$\sim$}}}

\def\be{\begin{equation}}
\def\ee{\end{equation}}
\def\bea{\begin{eqnarray}}
\def\eea{\end{eqnarray}}

\begin{document}

\title{FIRST ORDER PHASE TRANSITION OF THE Q-STATE POTTS MODEL 
IN TWO DIMENSIONS}

\author{H. ARISUE, K. TABATA}

\address{Osaka Prefectural College of Technology, 
Saiwai-cho, Neyagawa\\ Osaka 572, Japan\\
E-mail: arisue@las.osaka-pct.ac.jp} 


\maketitle\abstracts{ 
We have calculated the large-q series of the energy cumulants, the 
magnetization cumulants and the correlation length at the first order phase 
transition point both in the ordered and disordered phases for the $q$-state 
Potts model in two dimensions. The series enables us to estimate the numerical
values of the quantities more precisely by a factor of $10^2 - 10^4$ than 
the Monte Carlo simulations. From the large-q series of the eigenvalues 
of the transfer matrix, we also find that the excited states form a continuum 
spectrum and there is no particle state at the first order phase transition 
point.
}

\section{Introduction}

In many of the physical systems that exhibit the first order phase transition, 
the order of the transition changes to the second order by changing the 
parameter of the system. 
It is important to know how the quantities 
at the first order phase transition point diverge
when the parameter approaches the point at which the order of the transition 
changes.
The $q$-state Potts model\cite{Potts,Wu} in two dimensions gives
a good place to investigate this subject.
It exhibits the first order phase transition for $q>4$
and the second order one for $q\le 4$.
Many quantities of this model are known exactly 
at the phase transition point $\beta=\beta_t$ for $q>4$, including the 
latent heat\cite{Baxter1973} and the correlation 
length in the disordered phase,\cite{Klumper,Buffenoir,Borgs}
which vanishes and diverges, respectively, 
in the limit $q\rightarrow 4_+$. 
One the other hand other physically important quantities 
such as the specific heat,
the magnetic susceptibility, and the correlation length 
(in the ordered phase) at the transition point,
which also diverge as $q\rightarrow 4_+$, 
are not solved exactly.


Here we calculate the large-$q$ expansion series
of the energy cumulants including the specific heat 
and the magnetization cumulants including the magnetic susceptibility
in both the phases and the correlation length 
in the ordered phase at the transition point 
using the finite lattice method.\cite{Enting1977,Creutz,Arisue1984}
Obtained long series for the energy and magnetization cumulants
give the estimates of the quantities that are more precise by 
a factor of $10^{2}-10^{4}$ than the Monte Carlo simulations.
Especially its estimates are within an accuracy of $0.1\%$ 
at $q=5$, where the correlation length is as large as 
a few thousands of the lattice spacing.
Bhattacharya {\em et al.}\cite{Bhattacharya1994} made 
a stimulating conjecture
on the asymptotic behavior of the energy cumulants 
at the first order transition point; 
the relation between the energy cumulants and the correlation length
in their asymptotic behavior at the first order transition point 
for $q\rightarrow 4_+$ 
will be the same as their relation in the second order phase transition
for $q=4$ and $\beta\rightarrow \beta_t$, 
which is well known from their critical exponents.
The obtained series enables us to confirm the correctness 
of the conjecture.

As for the correlation length at the first order phase transition point,
the results of the Monte Carlo simulation\cite{Janke1994} and 
the density matrix renormalization group\cite{Igloi} indicate 
that the correlation lengths are very close to each other
in the ordered and disordered phases for $q \gtsim 10$. 
On the other hand, at the second order phase transition point($q\le4$)
their ratio is known to be $1/2$. 
It is interesting whether the ratio is exactly equal to unity, 
remains close to unity, or approaches $1/2$ when $q\to 4_+$.
To investigate it
we calculate the first few terms of the large-$q$ expansion 
for the eigenvalues of the transfer matrix and 
find that from the second largest to the $N$-th largest eigenvalues 
with $N$ the one-dimensional size of the lattice make a continuum spectrum 
in the thermodynamic limit both in 
the ordered and disordered phases.
We also calculate the long series of the second moment 
correlation length in both the phases, which serve to
investigate the behavior of the spectrum of the 
eigenvalues of the transfer matrix in the region of $q$ close to $4$.

\section{Finite lattice method}
Here we use the finite lattice 
method,\cite{Enting1977,Creutz,Arisue1984}
to generate the large-$q$ series 
for the Potts model,
instead of the standard diagrammatic method 
used by Bhattacharya {\em et al.}\cite{Bhattacharya1997} 
The finite lattice method can in general give longer series 
than those generated by the diagrammatic method 
especially in lower spatial dimensions. 
In the diagrammatic method, 
one has to list up all the relevant diagrams and count the number 
they appear.
In the finite lattice method we can skip this job and 
reduce the main work to the calculation of the expansion 
of the partition function for a series of finite size lattices,
which can be done using the straightforward site-by-site 
integration\cite{Enting1980,Bhanot1990} 
without the diagrammatic technique.
This method has been used mainly to 
generate the low- and high-temperature series in statistical systems 
and the strong coupling series in lattice gauge theory.
We note that this method  
is applicable to the series expansion 
with respect to any parameter other than the 
temperature or the coupling constant.
Using this method we calculated\cite{Arisue1999,Arisue2000}
 the series for the $n$-th energy cumulants
($n=0 - 6$) to $z^{23}$, $n$-th magnetization cumulants
($n=1 - 3$) to $z^{21}$,
 and second moment correlation length to $z^{19}$
with $z\equiv 1/\sqrt{q}$.

\section{Energy cumulants}
The latent heat ${\cal L}$ 
at the transition point are known to vanish 
at $q\rightarrow 4_+$ as
$$
{\cal L} \sim 3\pi x^{-1/2}\;,
$$
with $x=\exp{(\pi^2/2\theta)}$ and $2\cosh{\theta}=\sqrt{q}$.
Bhattacharya {\em et al.}'s conjecture says
that the $n$-th energy cumulants $F_{d,o}^{(n)}$ 
at the first order transition point $\beta=\beta_t$ will 
diverge for $q\rightarrow 4_+$ as
\begin{equation}
F_d^{(n)}, (-1)^n F_o^{(n)} \sim \alpha B^{n-2}
  \frac{\Gamma\left(n-\frac{4}{3}\right)}{\Gamma
    \left(\frac{2}{3}\right)}x^{3n/2-2}\;.\label{eq:asymp_form}
\end{equation}
The constants $\alpha$ and $B$ in Eq.(\ref{eq:asymp_form}) 
should be common to the ordered and disordered phases
from the duality relation for each $n$-th cumulants.

If this conjecture is true, 
the product $F^{(n)}{\cal L}^{3n-4}$ is a smooth function of $\theta$, 
so we can expect that the Pad\'e approximation of $F^{(n)}{\cal L}^p$ 
will give convergent result at $p=3n-4$.
It has been examined for the large-$q$ series obtained by the finite lattice 
methods for $n=2,\cdots,6$ both in the ordered and disordered phases, 
which in fact give quite convergent Pad\'e approximants 
for $p=3n-4$ and as $p$ leaves from this value the convergence of the 
approximants becomes bad rapidly. An example can be seen in Fig. 1
for n=2 in the disordered phase.
\begin{figure}[tb]
\caption{
Pad\'e approximants of $F_d^{(2)}{\cal L}^p$ at $q=4$ plotted versus $p$.         }
\hspace{2cm}
\epsfxsize=6.0cm
\epsffile{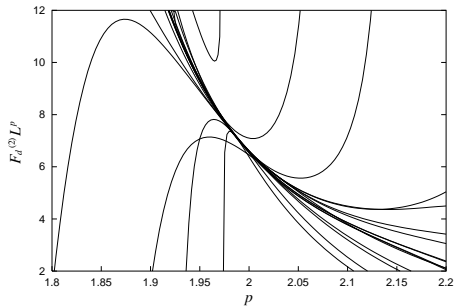}
\vspace{-0.5cm}
\end{figure}
We give in Table 1 the values of the 
specific heat $C=\beta_t^2F^{(2)}$ evaluated 
from these Pad\'e approximants for some values of $q$. 
These estimates are three or four orders of magnitude more precise than
(and consistent with)
the previous result for $q\ge 7$ 
from the large-$q$ expansion to order $z^{10}$
by Bhattacharya {\em et al.}\cite{Bhattacharya1997} and 
the result of the Monte Carlo simulations for $q=10,15,20$
carefully done by Janke and Kappler\cite{Janke} as in Table 2.
What should be emphasized is that we obtained the values of
the specific heat in the accuracy of about 0.1 percent
at $q=5$ where the correlation length is as large as 2500.
As for the asymptotic behavior of $F^{(n)}$ at $q\rightarrow 4_+$,
the Pad\'e data of $F_d^{(2)}/x$ and $F_o^{(2)}/x$  
have the errors of a few percent around $q=4$ and 
their behaviors are enough to convince us 
that the conjecture (\ref{eq:asymp_form}) is true for $n=2$ with
$\alpha = 0.073 \pm 0.002\;.$
Furthermore from the conjecture (\ref{eq:asymp_form}) the combination 
$\left\{{\Gamma\left(n-\frac{4}{3}\right)|F^{(n)}|}/
{\Gamma\left(\frac{2}{3}\right)F^{(2)}}\right\}^{\frac{1}{n-2}}
x^{-\frac{3}{2}}$ is expected 
to approach the constant $B$ for each $n(\ge 3)$, 
and in fact the Pad\'e data for every $n(=3,\cdots,6)$ gives 
$B=0.38\pm 0.05\;,$ 
which also gives strong support to the conjecture for $n\ge 3$.
\begin{table}[tb]
\caption{
The specific heat for some values of $q$.
The exact correlation length is also listed.
         }
\begin{center}
\begin{tabular}{|r|l|l|l|}
\hline
 $q$  & \multicolumn{1}{c|}{$C_d$}
         & \multicolumn{1}{c|}{$C_o$}
         & \multicolumn{1}{c|}{$\xi_d$(exact)} \\
\hline
$  5$ & 2889(2)         & 2886(3)         & 2512.2 \\ 
$  6$ & 205.93(3)       & 205.78(3)       & 158.9 \\ 
$  7$ & 68.738(2)       & 68.513(2)       & 48.1 \\ 
$  8$ & 36.9335(3)      & 36.6235(3)      & 23.9 \\ 
$  9$ & 24.58761(8)     & 24.20344(7)     & 14.9 \\ 
$ 10$ & 18.38543(2)     & 17.93780(2)     & 10.6 \\ 
$ 15$ & 8.6540358(4)    & 7.9964587(2)    & 4.2 \\ 
$ 20$ & 6.13215967(2)   & 5.36076877(1)   & 2.7 \\ 
\hline
\end{tabular}
\end{center}
\end{table}
\begin{table}[t]
\caption{
Comparison with the Monte Carlo simulations by Janke and Kappler(1997).
         }
\begin{center}
\begin{tabular}{|r|l|l|l|l|}
\hline
     &   & \multicolumn{1}{c|}{$q=10$}
        & \multicolumn{1}{c|}{$q=15$}
        & \multicolumn{1}{c|}{$q=20$} \\
\hline
$C_d$ & large-$q$     
         & $18.38543(2)$   &  $8.6540358(4)$ &  $6.13215967(2)$ \\ 
      & Monte Carlo
         & $18.44(4)$      &  $8.651(6)$     &  $6.133(4)$ \\ 
\hline
$C_o$ & large-$q$
         & $17.93780(2)$   &  $7.9964587(2)$ &  $5.36076877(1)$ \\ 
      & Monte Carlo
         & $18.0(1)$       &  $7.99(2)$      &  $5.361(9)$       \\ 
\hline
\end{tabular}
\end{center}
\end{table}

\section{Magnetization cumulants}
The behavior of the $n$-th magnetization cumulants $M^{(n)}$
for $\beta\to\beta_t$ at $q=4$ is well known as
$M_{d,o}^{(n)} \simeq A_{d,o}^{(n)} (\xi)^{\frac{15}{8}n-2}$
and parallel to the conjecture 
for the energy cumulants by Bhattacharya {\it et al.} 
we can make a conjecture that
\begin{equation}
M_{d,o}^{(n)} \sim \mu_{d,o}^{(n)} x^{\frac{15}{8}n-2}\;.\label{eq:asymp_m}
\end{equation}
in the limit $q\to 4_+$ with $\beta=\beta_t$. 
We have examined the Pad\'e approximation of $M^{(n)}{\cal{L}}^{p}$ 
for the large-$q$ series generated 
by the finite lattice method
for $n=2$ and $3$ both in the ordered and disordered phases, 
which in fact gives quite convergent Pad\'e approximants 
for $p=15n/4-4$
and as $p$ leaves from this value the convergence of the 
approximants becomes bad rapidly again. 
In Table 3 we present the resulting estimates of the magnetic susceptibility
$\chi_{d,o}=M_{d,o}^{(2)}$.
Our result is much more precise than the Monte Carlo simulation\cite{Janke} 
at least by a factor of 100 as in Table 4.
From the behavior of $M_{d,o}^{(n)}/x^{(15n/8-2)}$ 
we obtain the coefficients in Eq.(\ref{eq:asymp_m}) as 
$\mu_{d}^{(2)}=0.0020(2)$, $\mu_{o}^{(2)}=0.0016(1)$ 
and $\mu_{d}^{(3)}=7.4(5)\times 10^{-5}$, 
$\mu_{o}^{(3)}=7.9(2)\times 10^{-5}$. 
These convince us that the conjecture made for the magnetization cumulants 
is also true.

\begin{table}[t]
\caption{
The magnetic susceptibility for some values of $q$.
         }
\begin{center}
\begin{tabular}{|r|l|l|}
\hline
$q$  & \multicolumn{1}{c|}{$\chi_d$}
           & \multicolumn{1}{c|}{$\chi_o$} \\
\hline
$  5$ & $9.13(3)\times 10^4 $ & $9.01(3)\times 10^4  $ \\ 
$  6$ & $6.585(4)\times 10^2$ & $6.665(4)\times 10^2 $ \\ 
$  7$ & $70.54(1)           $ & $77.31(1)            $ \\ 
$  8$ & $19.359(1)          $ & $21.525(1)           $ \\ 
$  9$ & $8.0579(1)          $ & $9.0106(2)          $ \\ 
$ 10$ & $4.23276(2)         $ & $4.73823(4)         $ \\ 
$ 15$ & $0.7304214(1)       $ & $0.8056969(2)       $ \\ 
$ 20$ & $0.309365682(1)     $ & $0.33556421(1)      $ \\ 
\hline\end{tabular}
\end{center}
\end{table}
\begin{table}[t]
\caption{
Comparison with the Monte Carlo simulations 
by Janke and Kappler(1997).
         }
\begin{center}
\begin{tabular}{|r|l|l|l|l|}
\hline
   &     & \multicolumn{1}{c|}{$q=10$}
          & \multicolumn{1}{c|}{$q=15$}
        & \multicolumn{1}{c|}{$q=20$} \\
\hline
$M_d^{(2)}$ & large-$q$
            & $4.23276(2)$   &  $0.7304214(1)$ &  $0.309365682(1)$ \\ 
            & M.C.
            & $4.233(2)$     &  $0.73039(8)$   &  $0.30936(4)$     \\ 
\hline
$M_o^{(2)}$ & large-$q$
            & $4.73823(4)$   &  $0.8056969(2)$ &  $0.33556421(1)$  \\ 
            & M.C.
            & $4.74(4)$      &  $0.805(3)$     &  $0.3355(8)$      \\ 
\hline
\end{tabular}
\end{center}
\end{table}
\section{Exponential correlation length}
Next we investigate the exponential correlation length $\xi_{1,o}$ 
in the ordered phase. 
Here the exponential correlation length is defined 
by $\xi_1=\log{(\Lambda_1/\Lambda_0)}$ with $\Lambda_0$ and 
$\Lambda_1$ the largest and the second largest eigenvalues 
of the transfer matrix, respectively. (We have omitted the subscript '$1$'
in the previous sections)
There is an obstacle to extract the correction to the leading term of the 
large-$q$ expansion 
for the correlation length at the phase transition point 
from the correlation function $<{\cal O}(t){\cal O}(0)>_c$, 
since we know from the graphical expansion that it behaves like
\begin{eqnarray}
  <{\cal O}(t){\cal O}(0)>_c & \propto & z^t ( 1 + 2 z t^2 + \cdots ) 
                            \label{eq:graph}\\
                 &\ne& \exp{(- mt)}.\nonumber
\end{eqnarray}
for a large distance $t$.
Thus we will diagonalize the transfer matrix $T$ directly for large-$q$.
The eigenfunction for the largest eigenvalue $\Lambda_0$ 
in the leading order of $z$ is
$$
         |0> \equiv  |\underbrace{0\ 0\ 0\ ..... 0\ 0\ 0}_N>\;
$$
where all of the $N$ spin variables (each of which can take the value
of $s=0,1,\cdots,q-1$) are fixed to be zero,
with the element of the transfer matrix  $<0|T|0>$ $ = 1 + O(z^2)$
and the corresponding eigenvalue is $\Lambda_0 = 1 + O(z)$.
The eigenfunctions for the second largest eigenvalue
are 
$$
      |I> \equiv \frac{1}{\sqrt{N-I+1}}
                \sum |0\ ... 0\ \underbrace{e\ e\ e\ ... e\ e}_{I} 0\ ... 0>
$$
\vspace{-5mm}
$$
               (I=1,\cdots,N),
$$
$$
              |e> \equiv \frac{1}{\sqrt{q-1}} \sum_{s=1}^{q-1} |s>
$$
with the diagonal matrix elements
$$ <I|T|I> = z + O(z^2), $$
and the off-diagonal matrix elements starting from higher orders in $z$. 
The second largest eigenvalues of $T$ degenerate in the 
leading order with $\Lambda_i = z + O(z^{3/2})$.
The degeneracy of the eigenvalues 
of the first $N$ 'excited states' is the reason why the expansion series 
(\ref{eq:graph}) of the correlation function cannot be exponentiated into 
a single exponential term.
The off-diagonal matrix elements resolve the degeneracy 
with
\begin{eqnarray}
\Lambda_1/\Lambda_0 &=& z + 4 z^{3/2} + 6 z^{2} + O(z^3), \label{eq:gap}\\
                    & & \cdots                            \nonumber\\
\Lambda_N/\Lambda_0 &=& z - 4 z^{3/2} + 6 z^{2} + O(z^3).\nonumber
\end{eqnarray}
for $N \to \infty$.
These eigenvalues constitute a continuum spectrum. 
It appears to be kept in any higher order of $z$.
From Eq.(\ref{eq:gap}) we obtain
$$
1/\xi_{1,o} = - \log{z} - 4z^{1/2} + 2z +8/3z^{3/2}+ O(z^2). 
$$
This is the same as the large-$q$ expansion of $1/\xi_{1,d}$ given 
by Buffenoir and Wallon\cite{Buffenoir} to this order.

 In the disordered phase the situation is quite similar.
The eigenvalues of the transfer matrix for the first $N$ excited states 
constitute a continuum spectrum with their values exactly the same as 
in the ordered phase at least to the order of $z^{5/2}$.

The eigenvalues form the continuum spectrum just on the first order phase
transition point. Off the transition point, 
we can see that the spectrum is discrete with
      $\Lambda_i-\Lambda_{i+1} \sim O(\epsilon) $
         for $\sqrt{z}\ll  \epsilon \ll 1$ and
      $\Lambda_i-\Lambda_{i+1} \sim O(\epsilon^{2/3}) $
         for $\epsilon\ll \sqrt{z}  \ll 1$ 
where $\epsilon \equiv \beta/\beta_t -1$.

\section{Second moment correlation length}
Here we give the results of the large-$q$ expansion 
of the second moment correlation lengths $\xi_{2nd,o}$ 
in the ordered phase and $\xi_{2nd,d}$ in the disordered phase 
defined by 
$$
\xi_{2nd}^2=\frac{\mu_2}{2d\mu_0}\;,$$
where $\mu_2$ and $\mu_0$ are the second moment of the correlation function
and the magnetic susceptibility, respectively.
%
%
 The obtained expansion coefficients\cite{Arisue2000}
for the ordered and disordered phases
coincide with each other to order $z^{3}$ 
and differ from  each other in higher orders.
The ratio of the second moment correlation length in the ordered 
phase to that in the disordered phase is
estimated by the Pad\'e analysis to be not far from unity
even in the limit of $q \to 4$ ($\xi_{2nd,o}/\xi_{2nd,d}=0.930(3)$).
Another point is that
the ratio $\xi_{2nd,d}/\xi_{1,d}$ of the second moment correlation length 
to the exponential correlation length 
is much less than unity in the region of $q$ where the correlation length is
large enough. 
It approaches $0.51(2)$ for $q \to 4$. 
It is known that in the limit of the large 
correlation length, 
$$
       \frac{\xi_{2nd}^2}{\xi_{1}^2}
      \to \frac{\sum_{i=1}^{\infty} c_i^2 (\xi_i/\xi_1)^3}
               {\sum_{i=1}^{\infty} c_i^2 (\xi_i/\xi_1)}
      < 1,
$$ 
with $\xi_i\equiv -\log{(\Lambda_i/\Lambda_0)}$.
If the 'higher excited states' ($i=2,3,\cdots$) did not contribute so much, 
this ratio would be close to unity,
as in the case of the Ising model on the simple cubic lattice,
where $\xi_{2nd}/\xi_{1}=0.970(5)$ 
at the critical point.\cite{Caselle1999,Campostrini1998}
Our result implies that the contribution of the 'higher excited states' 
is large in the disordered phase of the Potts model in two dimensions
even when $q$ is close to $4$.
This strongly suggests that the eigenvalues of the 
transfer matrix for the first $N$ excited states in the disordered phase 
form the continuum spectrum not only in the large-$q$ region but also
when $q$ approaches $4$.

As for the exponential correlation length $\xi_{1,o}$ in the ordered phase
for $q\to 4$, it is difficult to calculate 
the eigenvalues of the transfer matrix in much higher orders. 
It is quite natural, however, to expect that 
the ratio $\xi_{1,o}/\xi_{1,d}$ would be close to unity
even in the limit of $q\to 4$.
The reason is the following.
If the ratio $\xi_{1,o}/\xi_{1,d}$ would be 
$1/2$ in the limit of $q\to 4$, 
which is the known ratio in the second order phase transition point($q\le4$),
then the ratio $\xi_{2nd,o}/\xi_{1,o}$ should 
be close to unity, which would imply that the higher excited states 
would not contribute so much to $\xi_{2nd,o}$ and 
the eigenvalue of the transfer matrix for the first excited state
would be separated from the higher excited states.
This scenario is not plausible, since as already mentioned in section 5
the continuum spectrum of the eigenvalues of the transfer matrix appears 
to be maintained in any high order in $z$.
In this case, we can expect that
the ratio $\xi_{2nd,o}/\xi_{1,o}$ would  be around $1/2$
as is the case in the disordered phase, resulting that 
$\xi_{1,o}/\xi_{1,d}$ is close to unity.

\section{Summary}
We generated the large-$q$ series for the energy and magnetization cumulants 
at the first order phase transition point 
of the two-dimensional $q$-state Potts model
in high orders using the finite lattice method. 
They gave very precise estimates of the cumulants for $q>4$
and confirmed the correctness of the Bhattacharya {\it et al.}'s 
conjecture that 
the relation between the cumulants and the correlation length 
for $q=4$ and $\beta\to\beta_t$ (the second order phase transition)
is kept in their asymptotic behavior for $q\to 4_+$ 
at $\beta=\beta_t$ (the first order transition point).
If this kind of relation is satisfied as the asymptotic behavior 
for the quantities at the first order phase transition point
in more general systems 
when the parameter of the system is varied 
to make the system close to the second order phase transition point,
it would serve as a good guide in investigating the system.

Further the large-$q$ expansion of the eigenvalues of the transfer matrix 
was calculated in the first 4 terms.
We found that
they have the same spectra of the eigenvalues of the transfer matrix 
in the ordered and disordered phases giving the same exponential
correlation length ($\xi_{1,o}=\xi_{1,d}$) to the order of $z^{3/2}$
and that the spectra are continuous in the thermodynamic limit.
We also calculated the large-$q$ expansion of the second moment 
correlation length in the ordered and disordered phases in high orders
and found that they differ from each other in higher orders than $z^{3}$, 
but that the ratio $\xi_{2nd,d}/\xi_{1,d} $ is not far from unity 
for all region of $q>4$.
We also found that $\xi_{2nd,d}/\xi_{d,1} $ is far from unity 
even in the limit of $q\to 4$. It receives significant contributions 
not only from the 'first excited state' but also 'higher excited states'
and this suggest strongly that the continuum spectrum would be maintained
(i.e. there would be no particle state) in the disordered phase.
From these results it is quite natural to expect that the 
exponential correlation length $\xi_{1,o}$ in the ordered phase is not 
far from that in the disordered phase even in the limit of $q\to 4$
and it is not plausible that their ratio approaches $1/2$ that
is their ratio in the second order phase transition point($q\le4$).


\end{document}